\documentclass{article}
\usepackage{waspaa21,amsmath,graphicx,url,times}
\usepackage{color}
\usepackage{graphicx,amsfonts,comment,array,multirow, cite, here}
\usepackage{subfig}
\usepackage{tikz}
\usetikzlibrary{intersections, calc, arrows.meta}
\usepackage{xcolor,colortbl}

\definecolor{Gray}{gray}{0.85}

\newcolumntype{a}{>{\columncolor{Gray}}c}

\usepackage{xcolor,url}

\def\thline{\noalign{\hrule height 0.8pt}}
\def\tthline{\noalign{\hrule height 1.4pt}}

\usepackage[normalem]{ulem}

\title{Training Speech Enhancement Systems with Noisy Speech Datasets}

\name{Koichi Saito$^{1}$,
      Stefan Uhlich$^{2}$,
      Giorgio Fabbro$^{2}$,
      Yuki Mitsufuji$^{2}$}
\address{$^1$ The University of Tokyo, Japan ~~~
         $^2$ Sony Corporation, R\&D Center, Germany/Japan}

\begin{document}
\ninept
\maketitle

\begin{sloppy}

\begin{abstract}
Recently, deep neural network (DNN)-based speech enhancement (SE) systems have been used with great success.
During training, such systems require clean speech data -- ideally, in large quantity with a variety of acoustic conditions, many different speaker characteristics and for a given sampling rate (e.g., 48kHz for fullband SE).
However, obtaining such clean speech data is not straightforward -- especially, if only considering publicly available datasets. 
At the same time, a lot of material for automatic speech recognition (ASR) with the desired acoustic/speaker/sampling rate characteristics is publicly available except being clean, i.e., it also contains background noise as this is even often desired in order to have ASR systems that are noise-robust.
Hence, using such data to train SE systems is not straightforward.
In this paper, we propose two improvements to train SE systems on noisy speech data.
First, we propose several modifications of the loss functions, which make them robust against noisy speech targets.
In particular, computing the median over the sample axis before averaging over time-frequency bins allows to use such data.
Furthermore, we propose a noise augmentation scheme for mixture-invariant training (MixIT), which allows using it also in such scenarios.
For our experiments, we use the Mozilla Common Voice dataset and we show that using our robust loss function improves PESQ by up to $0.19$ compared to a system trained in the traditional way.
Similarly, for MixIT we can see an improvement of up to $0.27$ in PESQ when using our proposed noise augmentation.
\end{abstract}

\begin{keywords}
Speech enhancement, Mozilla Common Voice, mixture-invariant training, deep neural networks
\end{keywords}

\section{Introduction}

\label{sec:intro}
Speech enhancement (SE) is a technique for extracting a clean speech signal from an observed signal containing speech and noise.
It is a fundamental technique for various audio applications including automatic speech recognition or voice calls. 
The recent development of SE has been built upon machine learning techniques using deep neural networks (DNNs)\cite{defossez2020real,koyama2020exploring,choi2020phaseaware,choi2021realtime,isik2020poconet}.

In supervised learning, a large amount of training data is required to obtain a well-performing system.
Most of the DNN-based SE systems require clean speech data for training.
Collecting such data is quite expensive as it requires recordings with well-controlled conditions.
Even if we can get clean speech data, the quantity and variety of acoustic conditions, and the speaker characteristics are often limited, especially given some constraints on the sampling rate (e.g., 48kHz for a fullband system).

To overcome such problems, researchers have tried to train DNN-based SE systems without clean speech data\cite{alamdari2020improving,fujimura2021noisytarget,MixIT,maciejewski2021training}.
The training strategy used in \cite{alamdari2020improving} was originally proposed in Noise2Noise~\cite{pmlr-v80-lehtinen18a}: it requires pairs of noisy signals consisting of the same speech data but different noise samples.
In image processing applications, we can obtain images of exactly the same target with different noise patterns using multiple exposures.
However, this is not the case for audio applications as it is in general impossible to observe multiple noisy signals with exactly the same speech signal and, hence, the multi-microphone approach of \cite{alamdari2020improving} has its limitations.
In \cite{maciejewski2021training, fujimura2021noisytarget}, the training strategy is the same as if clean speech data would be used, i.e., the training targets are the noisy speech. 
However, the experimental results in \cite{maciejewski2021training, fujimura2021noisytarget} showed that their proposed method could not achieve a better performance than a DNN trained with a smaller dataset of clean speech.
Finally, another approach is mixture-invariant training (MixIT) from~\cite{MixIT}. It proposes a loss function to train from a dataset of mixtures that can also be applied to SE with noisy targets. 
We performed some preliminary experiments which showed that the performance of an SE system trained with MixIT is poor as the noise in the speech target and the artificially added noise stem from different distributions: this conflicts with the ``independent mixture'' assumption of MixIT, causing a poor performance. In summary, no satisfactory solution for training with noisy speech data exists yet.

Hence, in this paper, we propose two improvements to train SE systems on noisy speech data.
First, we propose several modifications of the loss function, which make it robust against noisy speech by exploiting characteristics of time-frequency (T-F) bins. 
In particular, computing the median over the sample axis before averaging over the T-F bins allows to train from noisy speech data.
Second, we propose a noise augmentation scheme for MixIT, which allows using it also in such scenarios.

The main contributions of this paper are as follows. First, we analyze noise artefacts in the Mozilla Common Voice~\cite{MCV} dataset, which is a large, open-sourced and multi-lingual but noisy speech dataset. Second, we study several loss functions for SE systems with noisy speech data and propose a loss function which computes the median over the sample axis before averaging over the T-F bins. Finally, we propose a noise augmentation scheme for MixIT and confirm that this scheme allows to train with noisy speech data.
 
\section{Problem Definition and Related Work}
\label{sec:relatedworks}
 
\begin{figure*}[t]
\centering
\includegraphics[width=1.9\columnwidth]{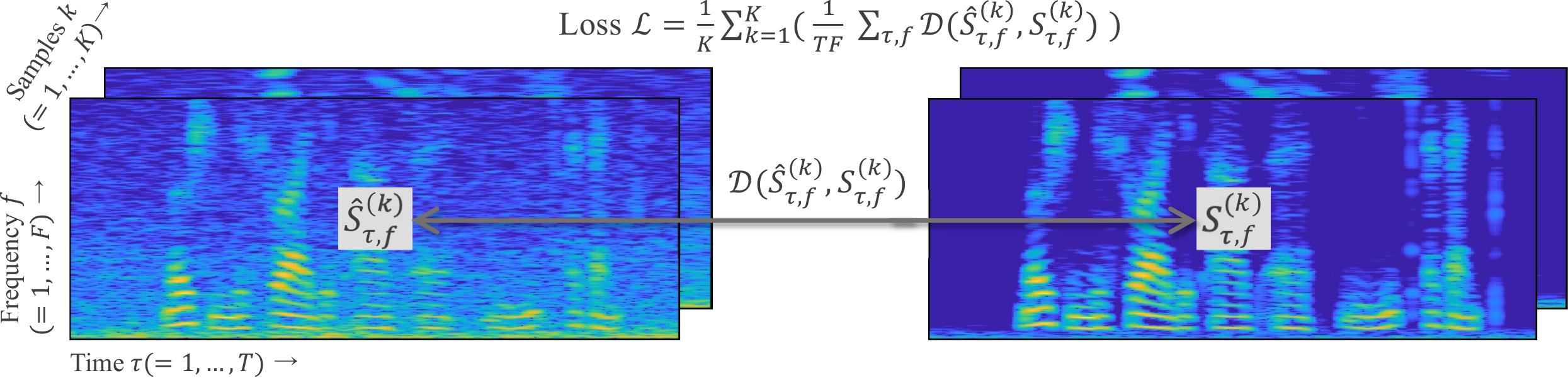}
\caption{Visualization of loss computation, the errors of individual T-F bins are averaged over time, frequency and sample dimension.}
 \label{fig:loss}
 \vspace{-1.0\baselineskip}  
\end{figure*}

\subsection{Problem settings of speech enhancement}
\label{ssec:problem_settings}
Let $\mathbf{x}\in\mathbb{R}^{T'}$ denote an observed mixture of target speech $\mathbf{s}\in\mathbb{R}^{T'}$ and noise $\mathbf{n}\in\mathbb{R}^{T'}$, i.e., $\mathbf{x} = \mathbf{s} + \mathbf{n}$ in the time domain.
The goal of SE is to reconstruct the speech signal $\mathbf{s}$ from $\mathbf{x}$.

Over the last years, numerous DNN-based SE systems have been proposed.
Many of those systems are based on T-F masking in the short-time Fourier transform (STFT) domain \cite{STFT1,STFT2,STFT3,STFT4,STFT5}.
Using T-F masking, the enhanced speech $\hat{\mathbf{s}}$ can be obtained by $\hat{\mathbf{s}} = \mathcal{F}^{-1}(\hat{\mathbf{S}}) = \mathcal{F}^{-1}(\mathcal{M}_\theta \odot \mathbf{X}))$,
where $\mathcal{F}^{-1}$ is the inverse STFT, $\mathcal{M}_\theta$ is a T-F mask, $\odot$ is the element-wise product, $\mathbf{X}$ is the T-F representation of $\mathbf{x}$ and $\hat{\mathbf{S}}$ is the T-F representation of $\hat{\mathbf{s}}$, respectively.
During training, the loss \begin{equation}
\vspace{-0.2\baselineskip}  
    \mathcal{L} = \frac{1}{K}\sum^{K}_{k=1}\left(\frac{1}{TF}\sum_{\tau,f}\mathcal{D}(\hat{S}^{(k)}_{\tau,f},S^{(k)}_{\tau,f})\right),
    \label{eq:loss}
\vspace{-0.2\baselineskip}
\end{equation}
between the enhanced signal $\hat{S}^{(k)}_{\tau,f}$ and target signal $S^{(k)}_{\tau,f}$ is minimized where
$k = 1,\dots, K$ denotes the (minibatch) sample index, $f = 1, \dots,F$ and $\tau = 1,\dots,T$ denote frequency and time indices, respectively. The difference between $\hat{S}^{(k)}_{\tau,f}$ and $S^{(k)}_{\tau,f}$ is measured by the distance $\mathcal{D}(.,.)$. Please see Fig.~\ref{fig:loss} for a visual depiction of the loss computation.

Most of the DNN-based systems require clean speech as targets $S_{\tau,f}^{(k)}$ during training.
However, as described in Sec.~\ref{sec:intro}, such data is often not available and also not straight-forward to generate.
At the same time, noisy speech data can be obtained more easily than clean data in a large amount and variety by using crowdsourcing.
For example, Mozilla Common Voice~\cite{MCV} is an open-source dataset that contains a large amount of speech data, variety of speakers and acoustic conditions (see Sec.~\ref{sec:mcv}).
Therefore, if we can train SE systems using noisy speech data which show a higher enhancement performance than using a small amount of clean speech data, we can expect further improvements for DNN-based SE systems.
\vspace{-0.0\baselineskip}  
\subsection{Mixture-invariant training (MixIT)}
\label{ssec:MixIT}
MixIT\cite{MixIT} is an unsupervised sound separation method, which requires only mixtures during training.
This method was proposed as a generalization of the permutation invariant training framework~\cite{PIT} to train directly from unsupervised mixtures.
MixIT can also be applied to SE as shown in the original paper and described in Fig.~\ref{fig:MixIT}.
When MixIT is applied in the T-F domain for SE, we input the mixture of two audio signals, noisy speech data and noise-only data, and we train by minimizing
\vspace{-0.2\baselineskip} 
\begin{multline}
    \mathcal{L}_{\rm{MixIT}}(\hat{\mathcal{X}}, \mathbf{X}, \mathbf{N})\\
    = \min(\mathcal{L}(\hat{\mathbf{X}}_{1}+\hat{\mathbf{X}}_{2}, \mathbf{X})
    +\mathcal{L}(\hat{\mathbf{X}}_{3}, \mathbf{N}), \\ \mathcal{L}(\hat{\mathbf{X}}_{1}+\hat{\mathbf{X}}_{3}, \mathbf{X})+\mathcal{L}( \hat{\mathbf{X}}_{2}, \mathbf{N})),
\label{eq:mixitloss}
\end{multline}
where $\hat{\mathcal{X}} = \{\hat{\mathbf{X}}_{1}, \hat{\mathbf{X}}_{2}, \hat{\mathbf{X}}_{3}\}$ denotes the three outputs of the DNN in the T-F domain, $\mathbf{X}$ the input noisy speech data in the T-F domain, $\mathbf{N}$ the noise-only data in the T-F domain, respectively.
By minimizing the loss \eqref{eq:mixitloss}, we find the best mixture permutation and we can expect that $\hat{\mathbf{X}}_{1}$ will contain enhanced speech data, while $\hat{\mathbf{X}}_{2}$ and $\hat{\mathbf{X}}_{3}$ will contain the noise. 
This scheme was used in~\cite{MixIT} for mixtures of speech with artificial noise.
However, in~\cite{maciejewski2021training} it has been reported that an SE system does not work well when using noisy speech data and noise signals as input and we will give an explanation of this phenomenon in Sec.~\ref{ssec:proposedmixit}.
\vspace{-0.5\baselineskip}  
\section{Mozilla Common Voice}
\vspace{-0.5\baselineskip}  
\label{sec:mcv}

\begin{table}[t]
\centering
\caption{Manual analysis of MCV \textit{valid} and \textit{invalid} for English. Classification is based on speech, noise and reverberation.}
\resizebox{\linewidth}{!}{
\begin{tabular}{c|c|c|c|c} \tthline
  Data & Click & \multirow{2}{*}{Reverberation} & \textit{valid}& \textit{invalid}\\ 
  attribution & noise & & (/300) &(/300)\\ \thline
  Clean speech & \multirow{2}{*}{w/o} & w/o & 60 & 37\\ 
  &  & w/ & 10 & 3\\ \cline{2-5}
  & \multirow{2}{*}{w/} & w/o & 82 & 32\\
  &  & w/ & 12 & 16\\ \cline{1-5}
  Noisy speech & \multirow{2}{*}{w/o} & w/o & 64 & 83 \\
   &  & w/ & 11 & 44\\ \cline{2-5}
   & \multirow{2}{*}{w/} & w/o & 52 & 27 \\
   & & w/ & 9 & 27\\\cline{1-5}
  Pure noise & - & - & 0 & 10\\
  Silence & - & - & 0 & 21\\ \thline
  \end{tabular}}
  \label{tab:mcv_analysis}
\end{table}

In this work, we used the Mozilla Common Voice (MCV) dataset for the training of our SE systems. In the following, we will describe this dataset in more detail.
MCV is an open-source, multi-lingual speech dataset.
The current release ``2020-12-11'' consists of $9,283$ recorded hours in $60$ languages, but the project is constantly growing, so more speakers and languages will be added in the future.
It is a crowd-sourced collection of voice clips featuring a wide variety of real-world recording environments and speaker characteristics.
However, due to the fact that the data collection is crowd-sourced, many of the recorded voice clips contain noise.

Submitted clips are sorted into two groups, MCV \textit{valid} and \textit{invalid}, according to certain rules.\footnote{\url{https://commonvoice.mozilla.org/en/about#how-it-works}}.
\textit{Valid} means that the speech samples can be used for automatic speech recognition (ASR) and \textit{invalid} means that the samples are not appropriate for ASR.
The MCV \textit{valid} subset contains only speech data and some of them contain background noises, reverberations and other sound.
MCV \textit{invalid} contains in addition silent data or noise-only data due to errors made by the contributors.

In order to get a better understanding of the data that is present in both subsets, we did a manual analysis where we used 300 randomly chosen samples from MCV \textit{valid} and \textit{invalid} in English. The results are summarized in Table~\ref{tab:mcv_analysis}. ``Click noise'' in Table~\ref{tab:mcv_analysis} refers to mouse/keyboard click sounds that are generated when contributors finish their recording of a sample.
``Reverberation'' refers to the reverberation caused by the recording environment.
MCV is expected to ease the problem of limited data as mentioned above, except containing noisy speech.
In this paper, we will use MCV \textit{valid} and \textit{invalid} as training targets to show the benefit of our methods.
\vspace{-0.0\baselineskip}  
\section{Proposed Method}
\label{sec:propoed}
\vspace{-0.0\baselineskip}  

\begin{figure}[t]
\centering
\includegraphics[width=0.8\columnwidth]{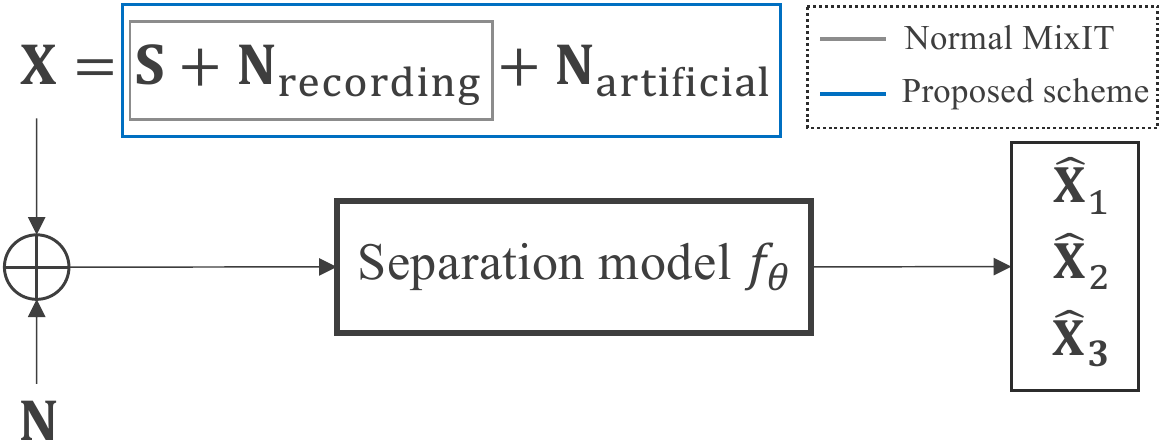}
\caption{Scheme of MixIT for SE and proposed noise augmentation. $\mathbf{S}$ denotes the clean speech signal, $\mathbf{N}_{\rm{recording}}$ the background noise signal during recording and $\mathbf{N}_{\rm{artificial}}$ is the noise signal which is added by our proposed augmentation, respectively.}
 \label{fig:MixIT}
\end{figure}

\begin{figure}[t]
\centering
\includegraphics[width=0.8\columnwidth]{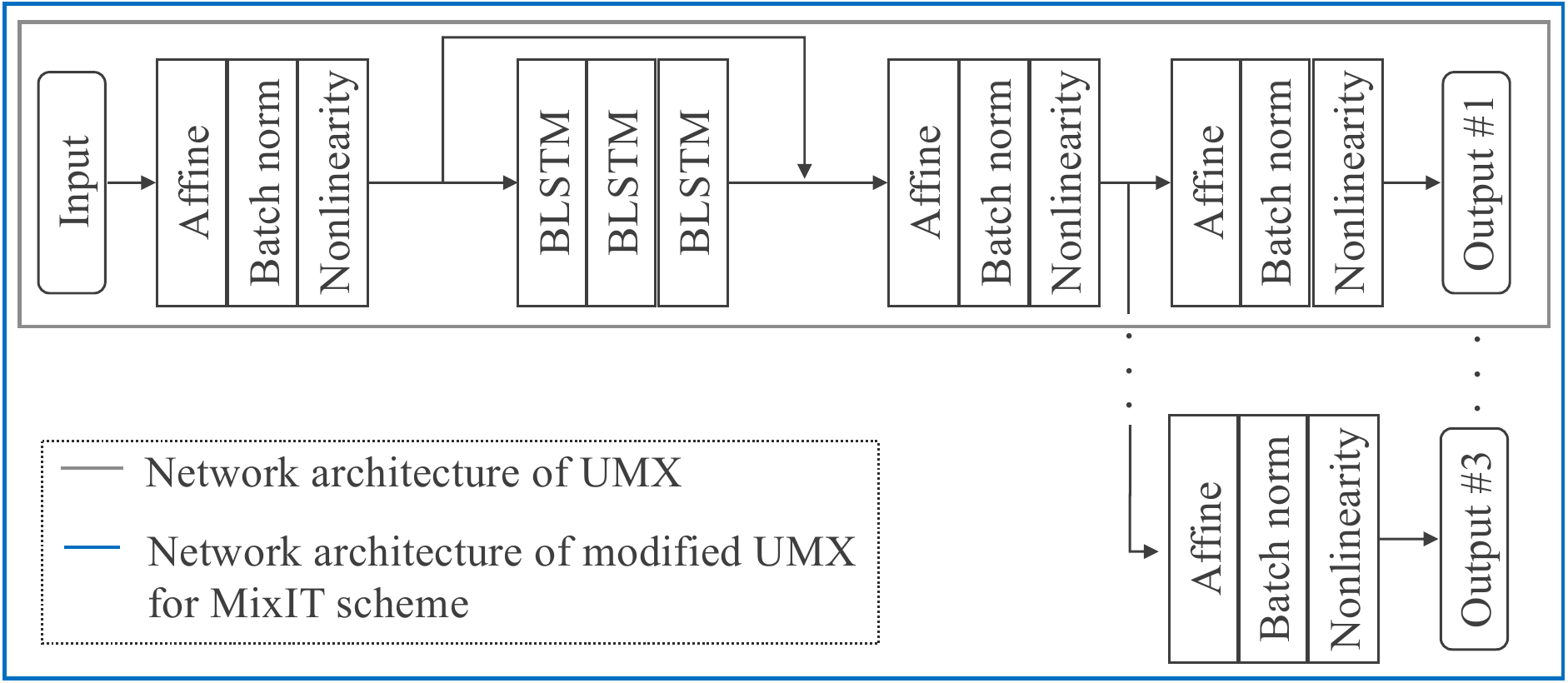}
\caption{Network architecture of Open-Unmix (UMX) and modified UMX for MixIT scheme.}
 \label{fig:archi_UMX}
\end{figure}

In order to train SE systems using noisy speech data, we propose two new methods.
First, we present in Sec.~\ref{ssec:lossfunction} several modifications of the loss \eqref{eq:loss} with the aim of making it robust against noisy speech targets.
Second, we propose in Sec.~\ref{ssec:proposedmixit} a noise augmentation scheme for MixIT, allowing to use it for our problem setting.
\vspace{-1.0\baselineskip}  
\subsection{Making the loss function robust}
\label{ssec:lossfunction}
To improve the SE performance when training with noisy speech data, we study the characteristics of noisy speech signals in the STFT domain and propose modifications to the mean squared error (MSE) loss.
The MSE is given by
\vspace{-0.2\baselineskip}  
\begin{multline}
    \mathcal{L}_{\rm{MSE}} = \frac{1}{K}\sum_{k=1}^{K}\left(\frac{1}{TF}\sum_{\tau,f}(\hat{S}^{(k)}_{\tau,f}-S^{(k)}_{\tau,f})^{2}\right)\\
    =\rm{MEAN}_{\textrm{Samples}}(\rm{MEAN}_{\textrm{T-Fbins}}\mathcal{D}(\bf{\hat{S}} - \bf{S}))
    \label{mse},
\end{multline}
where $K$ is the minibatch-size, $T$ is the number of time frames and $F$ is the number of frequency bins in the T-F domain.
We will refer in the following to the standard MSE as ``sample T-F bin mean''.
Similarly, the signal-to-distortion ratio (SDR) loss is given by
\vspace{-0.2\baselineskip}  
\begin{multline}
    \mathcal{L}_{\rm{SDR}} = -\frac{1}{K}\sum_{k=1}^{K}\left(\frac{1}{TF}\sum_{\tau,f}10\log_{10}\frac{(S^{(k)}_{\tau,f})^{2}}{(\hat{S}^{(k)}_{\tau,f}-S^{(k)}_{\tau,f})^{2}}\right).
    \label{sdr}
\end{multline}

Table~\ref{tab:proposedloss} lists our proposed loss functions.
We will now discuss each proposal and give the rationale behind it:
\vspace{-0.0\baselineskip}  
\begin{itemize}
\setlength{\itemsep}{-0.4mm} 
\item ``Sample median, T-F bin mean'': Compute mean over T-F bins and then median over samples. This is expected to train without using difficult samples for denoising (i.e., very noisy targets).
\item ``Sample mean, T-F bin median'': Compute median over T-F bins and then mean over sample. This is expected to train without using difficult T-F bins.
\item ``Sample mean, T-bin median, F-bin mean'': Compute mean over F-bins, median over T-bins and then mean over samples.
This is expected to train without using difficult time frames which, e.g., contain click noise.
\item ``T-F bin mean, sample median'': Compute median over samples and then mean over T-F bins. 
This loss is expected to train using only samples appropriate for enhancement and all T-F bins.
\item ``T-F bin mean, sample trimmed mean'': Compute mean over 25\% of samples from the minibatch with the smallest loss (for each T-F bin) and then mean over T-F bins.
Instead of ``T-F bin mean, sample median'', this loss can use more samples and thus, if there are more samples in the minibatch which we can trust, can yield better performance.
\end{itemize}
We will compare these loss functions in Sec.~\ref{ssec:results}.
\vspace{0.5\baselineskip}  

\begin{table}
\centering
\caption{Proposed loss functions. Subscripts \rm{F-bins} denote operation along $f$ axis, \rm{T-bins} denote operation along $\tau$ axis, and \rm{Samples} denote operation along sample axis $k$ in~\eqref{mse}.}
\resizebox{\linewidth}{!}{
\begin{tabular}{c|c} \tthline
  Loss name & Loss computation \\ \thline
  Sample median, T-F bin mean & $\rm{MEDIAN}_{\textrm{Samples}}(\rm{MEAN}_{\textrm{T-Fbins}}\mathcal{D}(\bf{\hat{S}} - \bf{S}))$\\
  Sample mean, T-F bin median & $\rm{MEAN}_{\textrm{Samples}}(\rm{MEDIAN}_{\textrm{T-Fbins}}\mathcal{D}(\bf{\hat{S}} - \bf{S}))$\\
  Sample mean, T-bin median, F-bin mean & $\rm{MEAN}_{\textrm{Samples}}(\rm{MEDIAN}_{\textrm{T-bins}}(\rm{MEAN}_{\textrm{F-bins}}\mathcal{D}(\bf{\hat{S}} - \bf{S}))$\\
  T-F bin mean, sample median & $\rm{MEAN}_{\textrm{T-Fbins}}(\rm{MEDIAN}_{\textrm{Samples}}\mathcal{D}(\bf{\hat{S}} - \bf{S}))$\\
  T-F bin mean, sample trimmed mean & $\rm{MEAN}_{\textrm{T-Fbins}}(\rm{TrimmedMEAN}_{\textrm{Samples}}\mathcal{D}(\bf{\hat{S}} - \bf{S}))$\\
  \tthline
  \end{tabular}
  }
  \label{tab:proposedloss}
\end{table}

\begin{table*}[t]
\centering
\caption{PESQ scores of proposed loss functions on VBD and DNS test dataset.} 
\scalebox{0.78}{
\begin{tabular}{c|aaacccaaa} \tthline
  Training dataset $\rightarrow$ & \multicolumn{3}{>{\columncolor[gray]{0.85}} c}{Trained w/ VBD} & \multicolumn{3}{c}{Trained w/ MCV \textit{valid}}& \multicolumn{3}{>{\columncolor[gray]{0.85}} c}{Trained w/ MCV \textit{invalid}} \\ 
   Test dataset $\rightarrow$ & VBD & DNS (w/ rev.) & DNS (w/o rev.) & VBD & DNS (w/ rev.) & DNS (w/o rev.) & VBD & DNS (w/ rev.) & DNS (w/o rev.) \\ \thline
   Noisy (Raw) & 1.97 & 1.82 & 1.58 & 1.97 & 1.82 & 1.58 & 1.97 & 1.82 & 1.58 \\ \thline
   ``Sample T-F bin mean'' (traditional MSE)& 2.26 & 1.52 & 1.87 & 2.11 & \bf{1.91} & 1.62 & 2.07 & 1.89 & 1.56 \\
  ``Sample median, T-F bin mean'' & 2.24 & 1.61 & 1.73 & \bf{2.15} & 1.84 & 1.55 & \bf{2.26} & 1.90 & 1.57 \\
  ``Sample mean, T-F bin median'' & 1.54 & 1.59 & 1.39 & 1.39 & 1.43 & 1.29 & 1.49 & 1.48 & 1.33 \\
  ``Sample mean, T-bin median, F-bin mean'' & 2.05 & \bf{1.84} & 1.63 & 1.94 & 1.70 & 1.40 & 1.95 & 1.79 & 1.39 \\
  ``T-F bin mean, sample median'' & 2.19 & 1.40 & 1.66 & 2.09 & 1.83 & 1.56 & 2.16 & \bf{1.94} & \bf{1.73} \\
  ``T-F bin mean, sample trimmed mean'' & 2.35 & 1.45 & 1.94 & \bf{2.15} & 1.90 & 1.60 & 2.14 & 1.89 & 1.57 \\
  SDR & \bf{2.53} & 1.42 & \bf{2.01} & 2.11 & 1.89 & \bf{1.63} & 2.17 & 1.90 & 1.65 \\\tthline
  \end{tabular}
  }
  \label{tab:lossfunction}
\end{table*}

\begin{table*}[t]
\centering
\caption{PESQ scores for different training schemes (traditional SE, MixIT, MixIT with proposed augmentation) on VBD and DNS test dataset. All schemes used the SDR loss \eqref{sdr}.}
\scalebox{0.87}{
\begin{tabular}{c|cccaaa} \tthline
  Training dataset $\rightarrow$ & \multicolumn{3}{c}{Trained w/ MCV \textit{valid}} & \multicolumn{3}{>{\columncolor[gray]{0.85}} c}{Trained w/ MCV \textit{invalid}} \\ 
   Test dataset $\rightarrow$ & VBD & DNS (w/ rev.) & DNS (w/o rev.) & VBD & DNS (w/ rev.) & DNS (w/o rev.) \\ \thline
   Traditional training scheme & 2.11 & \bf{1.89} & 1.63 & 2.17 & \bf{1.90} & 1.65 \\
   MixIT (w/o augmentation) & 2.00 & 1.85 & 1.49 & 1.99 & 1.85 & 1.46 \\
   \bf{MixIT (w/ augmentation)} & \bf{2.27} & \bf{1.89} & \bf{1.64} & \bf{2.26} & \bf{1.90} & \bf{1.66}\\\tthline
  \end{tabular}
  }
  \label{tab:mixitresults}
  
\end{table*}

\vspace{-0.0\baselineskip}  
\subsection{Noise augmentation scheme for MixIT}
\label{ssec:proposedmixit}

As mentioned in Sec.~\ref{ssec:MixIT}, our preliminary experiments with MixIT failed to produce good results.
In the experiment, we used speech from the MCV dataset as input target signal $\mathbf{X}$ and noise from the DEMAND\cite{DEMAND} dataset as noise signal $\mathbf{N}$ in \eqref{eq:mixitloss} and Fig.~\ref{fig:MixIT}.
We assume that the reason for the failure of the experiment is that the characteristics of background noises ${\mathbf{N}_{\rm{recording}}}$ from the MCV dataset are different than those ${\mathbf{N}}$ from the DEMAND dataset.
This mismatch led the SE systems to minimize the loss~\eqref{eq:mixitloss} without forcing it to separate the clean speech signal $\mathbf{S}$ from $\mathbf{N}_{\rm{recording}}$.
More precisely, without using noise augmentation, the DNN has no incentive to put speech into $\hat{\mathbf{X}}_1$ but could put it always into one of the other estimates $\hat{\mathbf{X}}_2$ or $\hat{\mathbf{X}}_3$.

To solve this problem, we propose a new noise augmentation scheme for MixIT, shown in Fig.~\ref{fig:MixIT}. MixIT is trained by inputing $\mathbf{X}$ as ${\mathbf{S}}+{\mathbf{N}_{\rm{recording}}}+{\mathbf{N}_{\rm{artificial}}}$ signals, which are created by sampling a noise signal ${\mathbf{N}_{\rm{artificial}}}$ from the same dataset as ${\mathbf{N}}$.
Augmenting with noise from the same distribution as $\mathbf{N}$, the DNN sees two kinds of noise with speech in the mixture and it will put the two noises into $\hat{\mathbf{X}}_2$ and $\hat{\mathbf{X}}_3$ and only $\hat{\mathbf{X}}_1$ is left for the speech. We will see the effectiveness of this augmentation in Sec.~\ref{ssec:results}.
\vspace{-0.0\baselineskip}  
\section{Experiments}
\label{sec:experiments}
\vspace{-0.0\baselineskip}  

\subsection{Experimental Settings}
\label{ssec:experimentalsettings}

In the experiments, we compared models trained separately with three datasets: Voice-Bank + DEMAND (VBD) dataset~\cite{VoiceBank,DEMAND}, MCV \textit{valid}, and MCV \textit{invalid}.
As test data, we used two different datasets: VBD and the dataset from the DNS challenge~\cite{DNS}.
In the VBD dataset, $11,572$ noisy/clean speech pairs are provided. We randomly extracted $300$ pairs from the training set and use them as validation set.
To match the number of training samples in VBD, we randomly extracted $11,572$ samples from both MCV \textit{valid} and MCV \textit{invalid}, respectively, and used $11,272$ of them for training and the remaining $300$ for validation.
For testing, VBD provides 824 noisy/clean speech pairs, while the DNS challenge dataset provides 300.
The DNS challenge provides both samples with and without reverberation. We used both for evaluation.
In order to use the VBD and DNS datasets for evaluation, we downsampled the MCV samples from $48$kHz to $16$kHz.
As evaluation metric we used the perceptual evaluation of speech quality (PESQ) measure~\cite{PESQ1}, as it is considered to be highly correlated with subjective quality~\cite{PESQ2}.

We experimented with the proposed loss functions and noise augmentation scheme using Open-Unmix (UMX)~\cite{UMX1,UMX2,uhlich2017improving}.
Fig.~\ref{fig:archi_UMX} shows the architecture of UMX.
For the MixIT experiments, we modified UMX and added two more output branches to obtain $\hat{\mathbf{X}}_1$, $\hat{\mathbf{X}}_2$, and $\hat{\mathbf{X}}_3$ which are needed for \eqref{eq:mixitloss} as shown in Fig.~\ref{fig:archi_UMX}.
Augmentation noise ${\mathbf{N}_{\rm{artificial}}}$ was taken from the DEMAND dataset.
We trained MixIT using the SDR loss \eqref{sdr}. 

Each model was trained with a batch size $16$ for $1,000$ epochs and we applied early stopping with a patience of $140$.
The sequence length of a minibatch (over time) was $1.0$ second.
The hidden size of the dense bottleneck layers was set to $512$. 
We used spectrograms as input, where the FFT size was $1,024$ samples with $1/4$ (= $256$ samples) overlap.
All trainings were conducted with the Adam optimizer~\cite{adam} and an initial learning rate of $0.001$. 

\vspace{-0.2\baselineskip}  
\subsection{Results}


\label{ssec:results}
Table~\ref{tab:lossfunction} shows the PESQ scores obtained with our proposed loss functions. 
The results on the VBD test set show that the ``sample median, T-F bin mean'' loss function led to the highest score for both networks trained with MCV \textit{valid} and \textit{invalid}.
Each of them improved the scores by $0.04$ and $0.19$ points compared to the traditional ``sample T-F bin mean'' MSE.
In particular, the score of the model trained on MCV \textit{invalid} was comparable to that of the model trained on VBD using the standard ``sample T-F bin mean'' MSE and evaluated on the VBD test set. When evaluating with DNS dry speech, ``sample median, T-F bin mean'' did not give comparable results. This is probably due to the fact that DEMAND was used in training, and the DNN overfitted it.
On the other hand, when we consider the mean scores of models trained on MCV \textit{valid} and \textit{invalid} and evaluated on DNS dry speech, both ``T-F bin mean, sample median'' and SDR show better performance than all the other losses.
This indicates that these loss functions are less likely to overfit the training data.
The evaluation on DNS (reverberant speech) shows that most of the scores when MCV \textit{valid} or MCV \textit{invalid} was used for training were higher when VBD was used.
This is because MCV \textit{valid} and \textit{invalid} include data with reverberation.

Table~\ref{tab:mixitresults} shows PESQ scores of our proposed noise augmentation for MixIT, compared to the traditional training scheme and the original MixIT approach.
Using our noise augmentation improves the scores by $0.27$ when training with MCV \textit{valid} and \textit{invalid} and evaluating on VBD, compared to the original MixIT.
Manual inspection of the output of the models (for all training datasets) showed that $\hat{\mathbf{X}}_1$, $\hat{\mathbf{X}}_2$, and $\hat{\mathbf{X}}_3$ all contained speech components, up to a certain extent. As expected, only $\hat{\mathbf{X}}_2$ and $\hat{\mathbf{X}}_3$ contain noise (see \eqref{eq:mixitloss}). On the other hand, when training without our augmentation strategy, $\hat{\mathbf{X}}_1$ and $\hat{\mathbf{X}}_3$ contained speech, while $\hat{\mathbf{X}}_2$ had almost no speech. 
This suggests that $\hat{\mathbf{X}}_1+\hat{\mathbf{X}}_3$ was optimized to be close to $\hat{\mathbf{X}}$, without any guarantee on $\hat{\mathbf{X}}_1$ having only speech.
These results match to our conjecture in Sec.~\ref{ssec:proposedmixit}.

The scores of MixIT with our augmentation were equal to or better than those of the traditional training scheme.
In particular, when performing the evaluation on VBD and training on MCV \textit{valid} or MCV \textit{invalid}, the score improved by $0.16$ and $0.09$, respectively.
This result shows the effectiveness of MixIT with our noise augmentation as we improved over the traditional training scheme.
The small improvements we see when evaluating on the DNS challenge dataset, compared to VBD, is possibly due to the fact that we used VBD as source for both $\mathbf{N}$ and ${\mathbf{N}_{\rm{artificial}}}$ when training. We want to address this in our future work.
\vspace{-0.2\baselineskip}  

\section{Conclusion}
\label{sec:conclusion}
\vspace{-0.0\baselineskip}  
In this paper, we tackled the problem of training DNN-based SE systems from noisy speech datasets. We proposed two improvements. First, we proposed several modifications of the loss function, which make it robust against noisy speech samples. Second, we proposed a noise augmentation scheme for MixIT.
Through an extensive experimental evaluation, we showed that the proposed ``sample median, T-F bin mean'' loss function led to higher PESQ scores than traditional ``sample T-F bin mean'' MSE loss, because the latter ignores the noise in the speech targets. Furthermore, for MixIT, we could see an improvement in the scores when using our proposed noise augmentation.

\vfill\pagebreak
\bibliographystyle{IEEEtran}
\bibliography{refs21}

\end{sloppy}
\end{document}